\documentclass[journal=jacsat,manuscript=article]{achemso}
\usepackage{hyperref}
\usepackage{color}
\usepackage{float}
\usepackage{dblfloatfix}
\usepackage{comment}
\usepackage{graphicx}
\usepackage{wrapfig}
\usepackage{amsmath}
\usepackage{adjustbox}
\usepackage{ragged2e}
\usepackage[svgnames,table]{xcolor}
\usepackage{array}
\usepackage{longtable}
\usepackage{changepage}
\usepackage{setspace}
\usepackage{hhline}
\usepackage{multicol}
\usepackage{tabto}
\usepackage{multirow}
\usepackage{makecell}
\usepackage{ulem}
\usepackage{cancel}
\usepackage{tikz}

\usepackage[version=4,arrows=pgf-filled,textfontname=sffamily,mathfontname=mathsf]{mhchem} 

\usepackage[version=4, arrows=pgf-filled,
 textfontname=sffamily,
 mathfontname=mathsf]{mhchem}

\author{Mehdi Ghasemi}
\email{mehdi.ghasemi@manchester.ac.uk}
\affiliation{Department of Chemical Engineering, The University of Manchester, Manchester M13 9PL, United Kingdom}

\author{Mohamad Ali Ghafari}
\affiliation{Independent Researcher, Tehran, Iran}

\author{Masoud Babaei}
\email{masoud.babaei@manchester.ac.uk}
\affiliation{Department of Chemical Engineering, The University of Manchester, Manchester M13 9PL, United Kingdom}

\author{Valentina Erastova}
\email{valentina.erastova@ed.ac.uk}
\affiliation{School of Chemistry, University of Edinburgh, Joseph Black Building, David Brewster Road, Edinburgh, EH9 3FJ, United Kingdom}

\title[An \textsf{achemso} demo]
  {Molecular Insights into Caprock Integrity of Subsurface Hydrogen Storage: Perspective on Hydrogen-induced Swelling and Mechanical Response}

\begin{document}
\newpage
\begin{abstract}
 The geological storage of hydrogen (\ce{H2}) requires reliable long-term caprock sealing, yet the nanoscale interactions between \ce{H2} and clay minerals remain critically underexplored despite their importance for storage security. This lack of understanding has limited the ability to predict mechanical stability and leakage risks in \ce {H2} storage formations. Using molecular simulations, this study investigates the swelling behavior and mechanical properties of sodium montmorillonite (Mt), a common smectite clay, under varying hydration states and interlayer \ce{H2} contents. Results show that \ce{H2} accelerates hydration-state transitions, narrows the stability window of crystalline swelling, and promotes asymmetric plume formation in confined interlayers. \ce{H2} alters cation and water coordination, thereby weakening Na$^+$–Mt electrostatic interactions and modulating H-bond networks at the interface and in the bulk. Mechanical analysis reveals pronounced anisotropy in Mt. In-plane stiffness is mainly governed by basal spacing expansion, whereas out-of-plane stiffness is highly sensitive to the initial presence of water or \ce{H2}, which weaken interlayer cohesion. Tensile and compressive strengths in the in-plane directions follow in-plane stiffness trends, while the out-of-plane tensile strength is governed by Mt–water H-bonds. The presence of \ce{H2} further promotes Mt sheets separation by disrupting nanoscale liquid bridges. Collectively, these results provide the first atomistic-scale evidence that intercalated \ce{H2} reshapes swelling energetics, elastic anisotropy, and failure pathways in Mt, highlighting critical nanoscale mechanisms that may compromise caprock integrity during underground \ce{H2} storage. 
\end{abstract}


\newpage
\section{Introduction}
Storage of hydrogen, \ce{H2}, at gigatonne capacities will form a central pillar of future sustainable energy systems, simultaneously mitigating CO$_2$ emissions and enabling the seasonal availability of green energy \cite{krevor2023subsurface,heinemann2021enabling}. As the global energy landscape transitions toward carbon neutrality, large-scale geological storage of \ce{H2} will be essential for maintaining the resilience, reliability, and adaptability of renewable energy infrastructures\cite{hassanpouryouzband2021offshore}. While underground \ce{H2} storage (UHS) stands as the most economically efficient solution for large-scale energy storage, persistent and complex challenges associated with long-term storage efficacy continue to overshadow its technological maturity compared to other storage systems, currently limiting the progress of UHS development\cite{edlmann2024challenging,ghafari2024wetting}. 

In this context, the failure of caprock sealing is a significant concern \cite{ugarte2022review}. Among the clay minerals in caprock, montmorillonite (Mt), with its unique properties including low permeability, high surface area, and swelling capacity, is a critical dual-role player, capable of enhancing sealing performance while also posing risks to caprock stability\cite{ghasemi2022molecular}. On one hand, Mt, like other clay minerals, can enhance caprock performance by forming dense, compact layers that act as effective seals, preventing fluid migration, with its fine-grained structure and fluid retention abilities further contributing to sealing efficiency \cite{kadoura2016molecular,cygan2012molecular}. However, on the other hand, these same properties can also negatively impact caprock stability under certain conditions. The high surface area of Mt facilitates processes such as ion exchange or gas intercalation, enhancing their interactions with fluids and making them susceptible to geomechanical and geochemical alterations, including structural changes and dissolution \cite{braid2024hydrogen,rahromostaqim2019molecular}.

The unique physical properties of \ce{H2}, such as its high diffusivity and lower density compared to CO$_2$, make it more likely to migrate toward the surface at a faster rate. Given these properties, \ce{H2}  should be stored deeper and in lower-permeability sites than CO$_2$ and CH$_4$ to ensure caprock confinement and prevent leakage. Our previous research revealed that \ce{H2}  can penetrate sub-nanopores as small as 0.5~nm in clay minerals, facilitated by its exceptional rotational flexibility compared to cushion gases like CO$_2$ \cite{kahzadvand2024risk}. \ce{H2}  molecules intercalated in clay-rich, water-saturated caprocks may become trapped due to \ce{H2}’s low solubility in water and the presence of complex, disconnected pore pathways. This raises a key question: how does trapped \ce{H2}  impact the caprock’s geomechanical properties, such as structural integrity, deformation, and long-term sealing?

To address this question, we focused on the swelling behavior of Mt, induced by \ce{H2}, and its subsequent impact on the geomechanical response. The swelling behavior of Mt has been extensively studied due to its sensitivity to hydration. Swelling proceeds via two primary mechanisms: (i) crystalline swelling, involving discrete interlayer water adsorption, and (ii) osmotic swelling at higher water uptake. Crystalline swelling occurs in stepwise transitions as water content increases, shifting from the dry state to a monolayer ($0\mathrm{W} \rightarrow 1\mathrm{W}$), and subsequently to a bilayer hydration state ($1\mathrm{W} \rightarrow 2\mathrm{W}$) \cite{sposito1999surface,cygan2004molecular}. These transitions are typically identified through changes in the basal spacing ($d_\text{001}$), which reflect interlayer expansion. While there is a well-established understanding of how CO$_2$ intercalates into clay minerals and its effects on swelling and mechanical properties \cite{rahromostaqim2019molecular,ho2019revealing,mendel2021interlayer,hunvik2022influence}, the unique physical properties of \ce{H2}  raise concerns about the direct applicability and transferability of CO$_2$-based knowledge to \ce{H2}  storage. These differences challenge the assumption that existing models and findings related to CO$_2$ storage can be fully adopted for \ce{H2}  storage \cite{kahzadvand2024risk}.

\subsection{This study}
In this study, we focus on the fundamental mechanism of \ce{H2}-induced swelling in Mt and perform comprehensive molecular-level simulations to investigate the impact of intercalated \ce{H2} under caprock-relevant thermodynamic conditions. Specifically, we examine how \ce{H2} intercalation affects the structural evolution and alters the mechanical behavior and macroscopic properties of the clay matrix. In doing so, we investigate the impact of interlayer-trapped \ce{H2} on Mt swelling under varying hydration states (1W and 2W), and elucidate how \ce{H2}-induced swelling influences both the elastic and failure behavior of Mt. Furthermore, we provide molecular-level insights into the mechanisms by which \ce{H2} modulates swelling and alters the mechanical properties of Mt. To the best of the authors’ knowledge, this is the first study to quantitatively evaluate, in a consecutive and integrated manner, the swelling behavior and mechanical property changes associated with \ce{H2} storage in Mt from a molecular perspective.

\newpage
\section{Methodology}
\subsection{System Preparation}

\begin{sloppypar} 
Montmorillonite (Mt), a well-studied member of the smectite clay mineral group, was selected as a representative layered silicate for this study. Its high sensitivity to water \cite{norrish1954crystalline}, low mechanical strength\cite{collettini2009fault}, and prevalence in reservoir caprock \cite{busch2008carbon} have made it a focal point in the evaluation of geological gas storage. Mt is particularly relevant due to its association with the development of the instability of clay-rich formations \cite{tetsuka2018effects}. Structurally, Mt features a three-layer (TOT) arrangement composed of an octahedral sheet (O-sheet) sandwiched by two tetrahedral sheets (T-sheet) \cite{cygan2021advances}. The chemical composition of the sodium-saturated Mt is 
\ce{Na_{0.75}[Si_{7.75}Al_{0.25}](Al_{3.5}Mg_{0.5})O_{20}(OH)4}. Isomorphic substitutions were introduced in accordance with Loewenstein’s rule, ensuring that no two substitutions occurred at adjacent sites. An extended unit cell (UC) slab was constructed and replicated $8 \times 4$ times in the \textit{x}-(41.28 nm) and \textit{y}-(35.86 nm) directions, respectively, to form a larger three-layer structure with varying interlayer spacings. As Mt in caprock varies with depth and can host different water saturation, and given our previous findings that \ce{H2} can potentially penetrate sub-nanopores of Mt and occupy these confined spaces \cite{kahzadvand2024risk}, we modeled the \ce{H2} gas phase under varying water saturation conditions. Accordingly, the \ce{H2} content considered varied from 0 to 1.25 per UC (0 to 40 molecules per system), whereas the water content varied between 0 to 10 molecules per UC (0 to 320 molecules per system). 
\end{sloppypar}

\subsection{Molecular Dynamics Simulations}

All MD simulations were performed using LAMMPS open-source software \cite{plimpton1995fast}. For modeling the clay mineral slab, the accurate and well-known force field CLAYFF \cite{cygan2004molecular} was used. This force field predominantly employs Lennard-Jones and Coulomb potentials to model non-bonded interactions within the clay slab, while bonded interactions are utilized to accurately simulate the hydroxyl structure. The interlayer water molecules were represented by the SPC/E model \cite{mark2001structure} that is compatible with CLAYFF force field \cite{cygan2004molecular}. A two-site model developed by Yang \textit{et al.} \cite{yang2005molecular} was also applied to \ce{H2} molecules, we previously confirmed its accuracy for modelling in clay systems \cite{ghasemi2022molecular}. Our simulation workflow consists of two main stages. First, we design the system to explore how \ce{H2} induces crystalline swelling. Then, we investigate how this swelling affects the mechanical properties of the system.

\subsubsection{Crystalline Swelling Process}

Initially, each clay system containing different numbers of \ce{H2} and water molecules underwent energy minimization, followed by \textit{NPT} simulations at $T = 373\,\mathrm{K}$ and $P = 10\,\mathrm{MPa}$ for 2~ns and 20~ns with time steps of 0.01~fs and 1~fs, respectively, to reach a stable interlayer distance. The final configurations were obtained through an additional 5~ns \textit{NVT} simulation prior to the evaluation of the mechanical properties. Energy minimization was carried out using the conjugate gradient method with convergence criteria of $10^{-6}$ kcal/mol for energy and $10^{-8}$ kcal/(mol·\AA{}) for stress. In the \textit{NPT} and \textit{NVT} simulations, the damping parameters of the Nosé–Hoover thermostat and barostat were set to 100 and 1000 times the integration time step, respectively. Periodic boundary conditions were applied in all three directions. A cut-off radius of 13~\AA{} was used for short-range nonbonded interactions, while long-range electrostatic interactions were computed using the Ewald summation method with an accuracy of $10^{-5}$.  The comparison of Mt swelling values with experimental data and previous MD simulation studies is provided in the Supporting Information (SI), Table S1, and details of the energy- and trajectory-based analyses used to evaluate the impact of \ce{H2} on Mt swelling are also presented therein.

\subsubsection{Mechanical Properties}

The mechanical response of the Mt systems was quantified by analyzing their stress and elastic tensors. The stress tensor, $\sigma_{ij}$, is obtained from the virial expression:  

\begin{equation}
\sigma_{ij} = - \frac{1}{V_0} \left[ \sum_{i=1}^N m_i \, (\mathbf{v}_i \mathbf{v}_i^T) + \sum_{i<j} \mathbf{r}_{ij} \mathbf{f}_{ij}^T \right],
\end{equation}

where $i$ and $j$ run over all particles ($1 \leq i,j \leq N$). In this expression, 
$m_i$, $\mathbf{v}_i$, $\mathbf{r}_{ij}$, and $\mathbf{f}_{ij}$ denote the mass, velocity, displacement vector, and interatomic force acting on particle $i$ due to particle $j$, respectively. To describe the elastic behavior, the fourth-rank elastic stiffness tensor $C_{ijkl}$ is introduced. This tensor is defined as the second derivative of the potential energy $u$ with respect to the strain components $\varepsilon_{ij}$ and $\varepsilon_{kl}$, normalized by the reference cell volume $V_0$ \cite{gale2003general}:  

\begin{equation}
C_{ijkl} = \frac{1}{V_0} \frac{\partial^2 u}{\partial \varepsilon_{ij} \, \partial \varepsilon_{kl}}, 
\quad i,j,k,l \in \{1,2,3\}.
\end{equation}

Here, the indices $i$, $j$, $k$, and $l$ correspond to Cartesian directions.  
The stress and strain tensors are related through Hooke’s law, which for small deformations can be expressed as:  

\begin{equation}
\sigma_{ij} = C_{ijkl} \, \varepsilon_{kl}, 
\quad i,j,k,l \in \{1,2,3\}.
\end{equation}

Details of the stress tensor components are provided in Figure S1, SI. Although this tensorial form is rigorous, it is not practical for computational analysis. Therefore, Hooke’s law is often reformulated in the Voigt notation, which replaces double indices $(ij)$ by single indices $(1$ to $6)$. 
The mapping is given as $(11)\rightarrow 1$, $(22)\rightarrow 2$, $(33)\rightarrow 3$, $(23)$ or $(32)\rightarrow 4$, $(13)$ or $(31)\rightarrow 5$, and $(12)$ or $(21)\rightarrow 6$ \cite{voigt1910lehrbuch}. With this convention, the fourth-rank stiffness tensor is reduced to a $6 \times 6$ matrix form. Due to the orthotropic symmetry of clay platelets, only nine independent stiffness coefficients are needed to characterize the nanoscale elastic response \cite{ebrahimi2012nanoscale}. These coefficients can be grouped into two categories: in-plane (\textit{C$_{11}$}, \textit{C$_{22}$}, \textit{C$_{12}$}, \textit{C$_{66}$}) and out-of-plane ($C_{13}$, $C_{23}$, $C_{33}$, $C_{44}$, $C_{55}$). This distinction reflects the anisotropic nature of layered materials, where in-plane bonding differs significantly from interlayer interactions. Under the Voigt notation, Hooke’s law becomes:  

\begin{equation} \label{eq:stress_strain}
\begin{bmatrix}
\sigma_1 \\ \sigma_2 \\ \sigma_3 \\ \sigma_4 \\ \sigma_5 \\ \sigma_6
\end{bmatrix}
=
\begin{bmatrix}
C_{11} & C_{12} & C_{13} & 0 & 0 & 0 \\
 & C_{22} & C_{23} & 0 & 0 & 0 \\
& & C_{33} & 0 & 0 & 0 \\
& \text{sym} & & C_{44} & 0 & 0 \\
& & & & C_{55} & 0 \\
& & & & & C_{66}
\end{bmatrix}
\cdot
\begin{bmatrix}
\varepsilon_1 \\ \varepsilon_2 \\ \varepsilon_3 \\ \varepsilon_4 \\ \varepsilon_5 \\ \varepsilon_6
\end{bmatrix}
\end{equation}

where ``sym'' indicates symmetric terms that are not explicitly written. This compact form facilitates both computational evaluation and physical interpretation of the elastic properties of Mt. Further details on the conceptual definition of stiffness coefficients are provided in Table S2, SI.

Several methods exist to calculate the stiffness coefficients, including the stress-fluctuation \cite{squire1969isothermal,van2005isothermal}, strain-fluctuation \cite{parrinello1982strain,van2003improved}, and constant-strain methods \cite{theodorou1986atomistic,mazo2008molecular}. In this study, the constant-strain method was employed due to its proven reliability \cite{clavier2023computation,clavier2017computation}. In this method, the system is first energy-minimized and then systematically deformed along 12 independent strain directions. The strain in each direction varied from $-0.01$ (compression) to $+0.01$ (tension), with a step size of $10^{-4}$. The reported stiffness coefficients correspond to the average tensile and compressive responses in each principal direction.

To further analyze the mechanical response of Mt under tension and compression along the \textit{x}-, \textit{y}-, and \textit{z}-directions, a second set of simulations was performed under uniaxial strain loading. In this setup, the simulation box was stretched or compressed in one direction, while pressure and temperature were maintained in the other two directions. The temperature was fixed at $373\,\mathrm{K}$ in all directions, with zero pressure in the \textit{x}- and \textit{y}-directions and $10\,\mathrm{MPa}$ in the \textit{z}-direction, regulated using the Nosé–Hoover thermostat (damping parameter: $100\,\mathrm{fs}$) and barostat (damping parameter: $1000\,\mathrm{fs}$). Pressure was controlled in the two unstrained directions, while the third was deformed at a prescribed strain rate of $10^{-7}\,\text{fs}^{-1}$, consistent with previous studies demonstrating insensitivity of the results to this value \cite{wei2022effect,wei2023atomistic,niu2025multiscale}. The maximum applied strain reached 0.4. A detailed validation of the simulation framework—including comparisons with MD and experimental data—is provided in the SI, Table S1 for swelling behavior and Tables S3–S5 for mechanical behavior, confirming the accuracy and robustness of the employed methods. Note that a detailed description of all analyses adopted in this work is provided in the SI.


\newpage
\section{Results and Discussion}

To investigate the swelling behavior and mechanical properties of Mt under varying water and \ce{H2} contents, we focused on interlayer dynamics and structural responses. By analyzing basal spacing, hydration energy, intercalation mechanisms, and both elastic and failure behaviors, we elucidate how interlayer water and \ce{H2} affect Mt’s nanoscale interactions and macroscopic properties, offering insights relevant to geological and engineering applications.

\subsection{Impact of \ce{H2} on the Basal Spacing} 

\begin{figure}[H]
    \centering
    \includegraphics[width=1\linewidth]{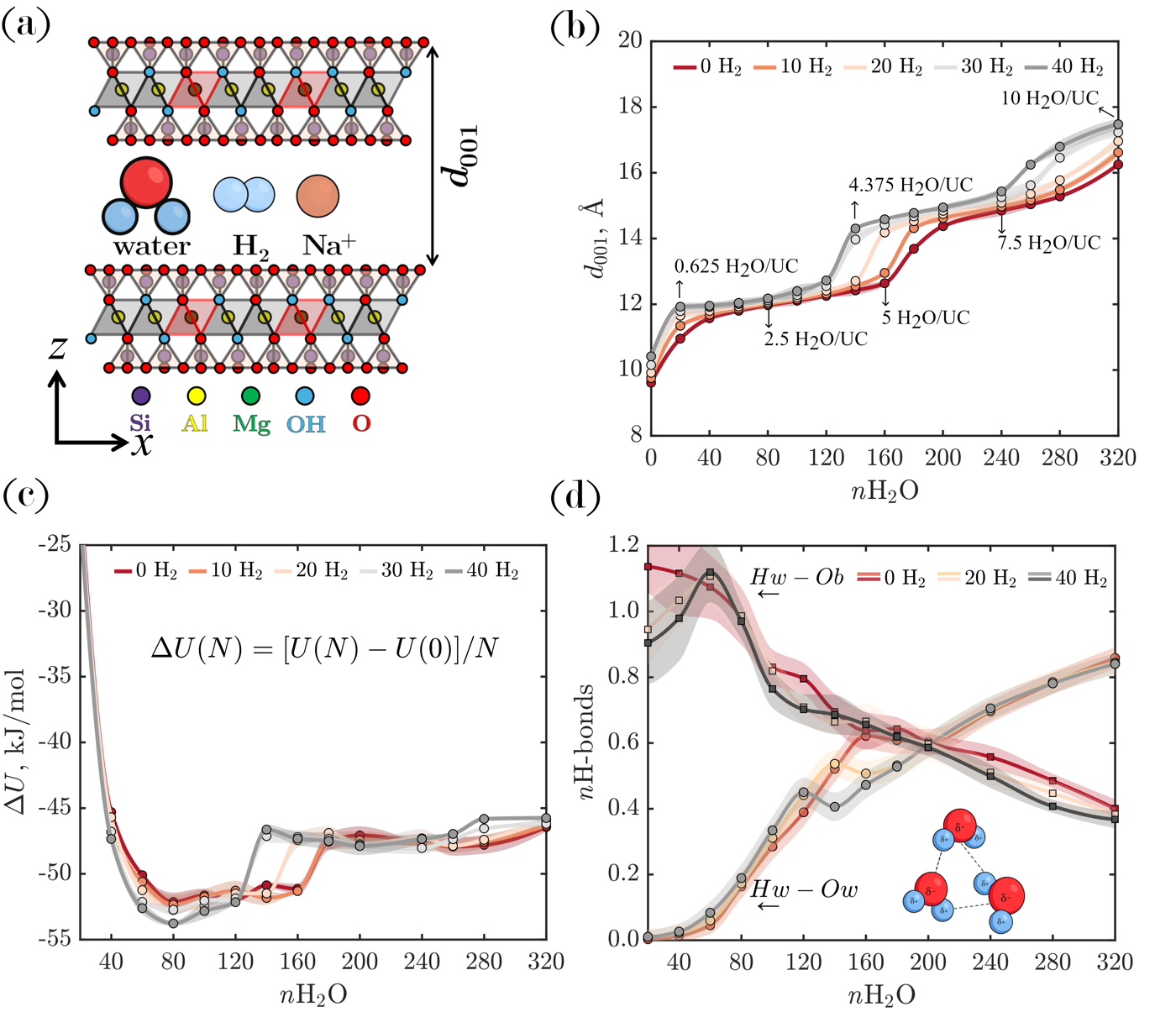}
        \caption{(a) Schematic of the system containing water, \ce{H2}, and cations, with basal spacing, $d_{001}$, defined as the distance from the top of one clay layer to the top of the next. (b–d) Variation of key structural and energetic properties of montmorillonite (Mt) as a function of water content (0 to 320 per 32 UCs)  under different \ce{H2}  loadings (0 to 40 \ce{H2} molecules per system): (b) $d_{001}$; (c) average hydration energy ($\Delta U(N)$); (d) average number of H-bonds per water molecule, including water--surface oxygen of Mt (\textit{Hw--Ob}) and water--water (\textit{Hw--Ow}). For clarity, panel (d) includes results for systems with 0, 20, and 40 \ce{H2} molecules only. The shaded region denotes the mean value $\pm$ standard deviation calculated over the equilibrium time window.}
    \label{fig:Swelling-Energy-Hbonds}
\end{figure}

Figure~\ref{fig:Swelling-Energy-Hbonds}a illustrates the definition of the basal spacing, $d_{001}$, while its variation as a function of interlayer hydration and \ce{H2} is shown in Figure~\ref{fig:Swelling-Energy-Hbonds}b. In the absence of both water \ce{H2}, the $d_{001}$ of anhydrous Mt is 9.7~\AA. With the increase of interlayer water to 1.25 water molecules per unit cell (UC) (40 \ce{H2O} confined within 32 UCs, represented by red line in Figure \ref{fig:Swelling-Energy-Hbonds}b), this spacing increases to 11.5~\AA\ due to formation of a monolayer (1W) hydration state. Further water uptake to 5 \ce{H2O}/UC slightly expands $d_{001}$ to 12.6~\AA, remaining in the stable 1W state. Above 5 \ce{H2O}/UC, the $d_{001}$ expands sharply to approximately 14.0~\AA, passing through a transition region with a steep slope in the $d_{001}$ curve and reaching the next configuration, in which water molecules arrange into a double layer (2W) hydration state. This state remains stable up to approximately 10 \ce{H2O}/UC, with the $d_{001}$ of $\sim$ 16~\AA\, after which triple layer (3W) configuration takes over. The spacing in all states of Mt is consistent with values reported in the literature (see Table S1) \cite{holmboe2014molecular,berend1995mechanism}. 

Under anhydrous conditions, upon exposure to \ce{H2}, the Mt interlayer expands to accommodate gas molecules by modifying the balance of intermolecular forces \cite{li2020internal} and enhancing interlayer repulsion (Figure  \ref{fig:Swelling-Energy-Hbonds}b, red line). In the presence of water (Figure \ref{fig:Swelling-Energy-Hbonds}b, pink to grey lines) , the impact of \ce{H2} on the interlayer spacing is minimal during stable 1W regions, yet it is prevalent during the transition between these hydration states. This behavior can be described as an `expansion--filling--saturation'' process \cite{teich2015swelling, li2020internal}. Therefore, \ce{H2} primarily affects anhydrous conditions and transition regions, accelerating the onset of interlayer expansion to the next stable hydration state. For instance, the 1W$\rightarrow$2W transition without \ce{H2} occurs when interlayer hydration reaches $\sim$ 5 \ce{H2O}/UC (160 water molecules), yet with \ce{H2} this transition occurs as early as 3.75 \ce{H2O}/UC (120 water molecules). This effect scales with \ce{H2} content, with 20 \ce{H2}  molecules showing intermediate behavior. In the 2W $\rightarrow$ 3W transition, \ce{H2} further increases the interlayer spacing and accelerates the transition, with a cumulative impact on swelling. Interestingly, unlike the 1W stable region—where the presence of \ce{H2} had minimal structural impact—the 2W stable region exhibits a more pronounced difference in $d_{001}$ depending on the \ce{H2} content.

Figure \ref{fig:Swelling-Energy-Hbonds}c shows the changes in average hydration energy, $\Delta U(N)$, with increasing \ce{H2} content. During the 1W stage, $\Delta U(N)$ of Mt decreases rapidly and eventually falls below the internal energy of pure SPC/E water model, which is around -41.5 kJ/mol\cite{berendsen1987missing}, indicating that the system would readily rehydrate if water were available. While all systems initially follow a similar trend, the presence of \ce{H2} leads to a greater reduction in $\Delta U(N)$, resulting in an even deeper global energy minimum. In the case of Mt (Figure \ref{fig:Swelling-Energy-Hbonds}c, red line) without \ce{H2}, the minimum energy region is relatively broad (spanning from 1.85 to 5 \ce{H2O}/UC), indicating energetic favorability across a wide range of hydration. In contrast, while the systems containing \ce{H2} exhibit lower overall potential energy, the well is much narrower (between 1.85 and 3.75 \ce{H2O}/UC), implying that their stability is energetically favorable only within a limited hydration range. In other words, \ce{H2}  facilitates a lower-energy pathway for enhanced swelling. This suggests that \ce{H2}-containing systems may exhibit reduced structural resilience when exposed to fluctuations in water content.

\subsection{Interlayer Molecular Interactions Driving Swelling}

\begin{figure}[H]
    \centering
    \includegraphics[width=0.95\linewidth]{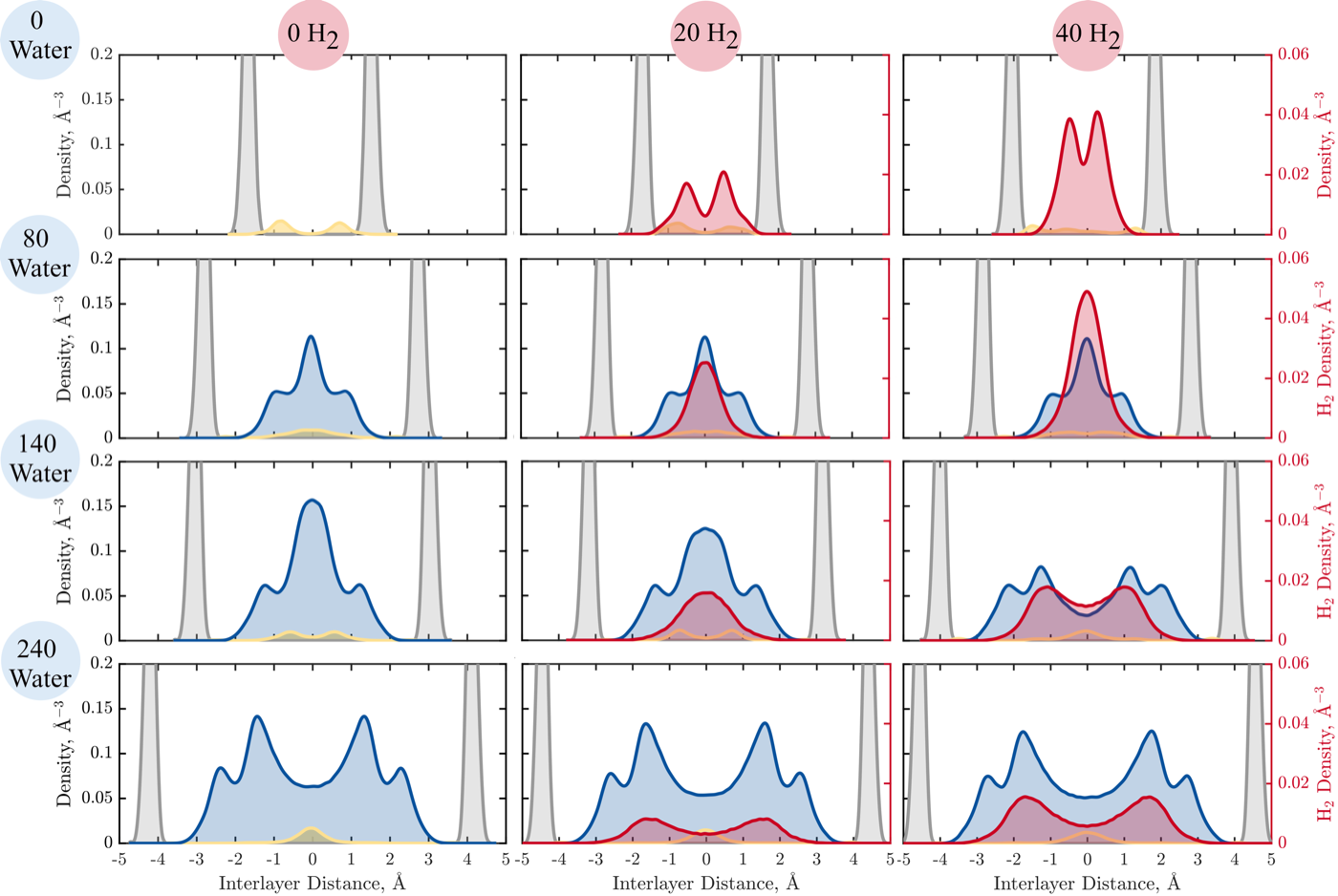}
    \caption{Atomic density profiles of \ce{H2O} (blue), \ce{H2} (red), \ce{Na+} (yellow), and Ob atoms of the clay surface (gray) at four different water contents representing dry clay (0 water; 0 \ce{H2O}/UC), a single water hydration layer (80 water; 2.5 \ce{H2O}/UC), a transition region (140 water; 4.375 \ce{H2O}/UC), and two water hydration layers (240 water; 7.5 \ce{H2O}/UC). Each profile is shown for three different \ce{H2} contents, highlighting the spatial distribution of species relative to the clay surface under varying hydration and \ce{H2} conditions.}
    \label{fig:1D-density}
\end{figure}

To understand the molecular interactions responsible for the observed swelling behavior, we examine how \ce{H2}  molecules influence the behavior of other species (cations and water) within the Mt interlayer. Figure \ref{fig:1D-density} shows linear atomic density profiles across the simulation box, perpendicular to the clay layer (along \textit{z}-axis). 

In anhydrous Mt, charge-balancing Na$^+$ ions form inner-sphere surface complexes (ISSCs), directly coordinated to the clay surfaces (Figure \ref{fig:1D-density}, yellow line, showing two distinct peaks at \(\sim 9\)~\AA{} from the peaks corresponding to surface oxygen, O\textsubscript{b}). Formation of ISSCs at distances less than 1.16~\AA{}, the ionic radius of the cation, indicates considerably closer coordination over the surface’s hexagonal cavities \cite{boek1995monte,pollak2024modeling}. In contrast, when \ce{H2} is absent, with increased hydration, Na$^+$ ion becomes hydrated and moves toward the interlayer center, positioning at 3.1 ~\AA{} from the surface as an outer-sphere surface complex (OSSC). At low hydration levels, this occurs for only a few \ce{Na+}, while others remain near the surface, determined by the available water to fully hydrate each cation \cite{holmboe2014molecular}.
This progression underscores the central role of Na$^+$ hydration energy in driving the swelling behavior of the clay \cite{anderson2010clay}. Presence of \ce{H2} at very low water content (0.625 \ce{H2O}/UC) shifts Na$^+$ slightly toward the clay slabs. However, as hydration increases, \ce{Na+} is gradually moved into the OSSC. More intensely, under 2W hydration (\textit{e.g.}, 7.5 \ce{H2O}/UC), when a large amount of \ce{H2} is present (\textit{e.g.}, 40 \ce{H2}), it positions itself between the clay surface and the cation, weakening the electrostatic interaction and allowing some Na$^+$ ions to drift into the geometric center of the interlayer space. This observation is further supported by the radial distribution function (RDF) between Na$^+$ and surface oxygen atoms (Na–O$_{\text{b}}$), which indicates a shift toward configurations characteristic of increased hydration (Figure S3, SI). Therefore, from the viewpoint of cation hydration, \ce{H2} facilitates the relocation of Na$^+$ into a hydrated state before the formation of a stable 1W layer. Once the 1W structure is established, however, \ce{H2} has little influence on the arrangement of the interlayer components. In the 2W state, \ce{H2} tends to shift Na$^+$ ions toward the center of the interlayer, indicating a greater contribution of cation hydration to the observed swelling.

\begin{figure}[H]
    \centering
    \includegraphics[width=0.95\linewidth]{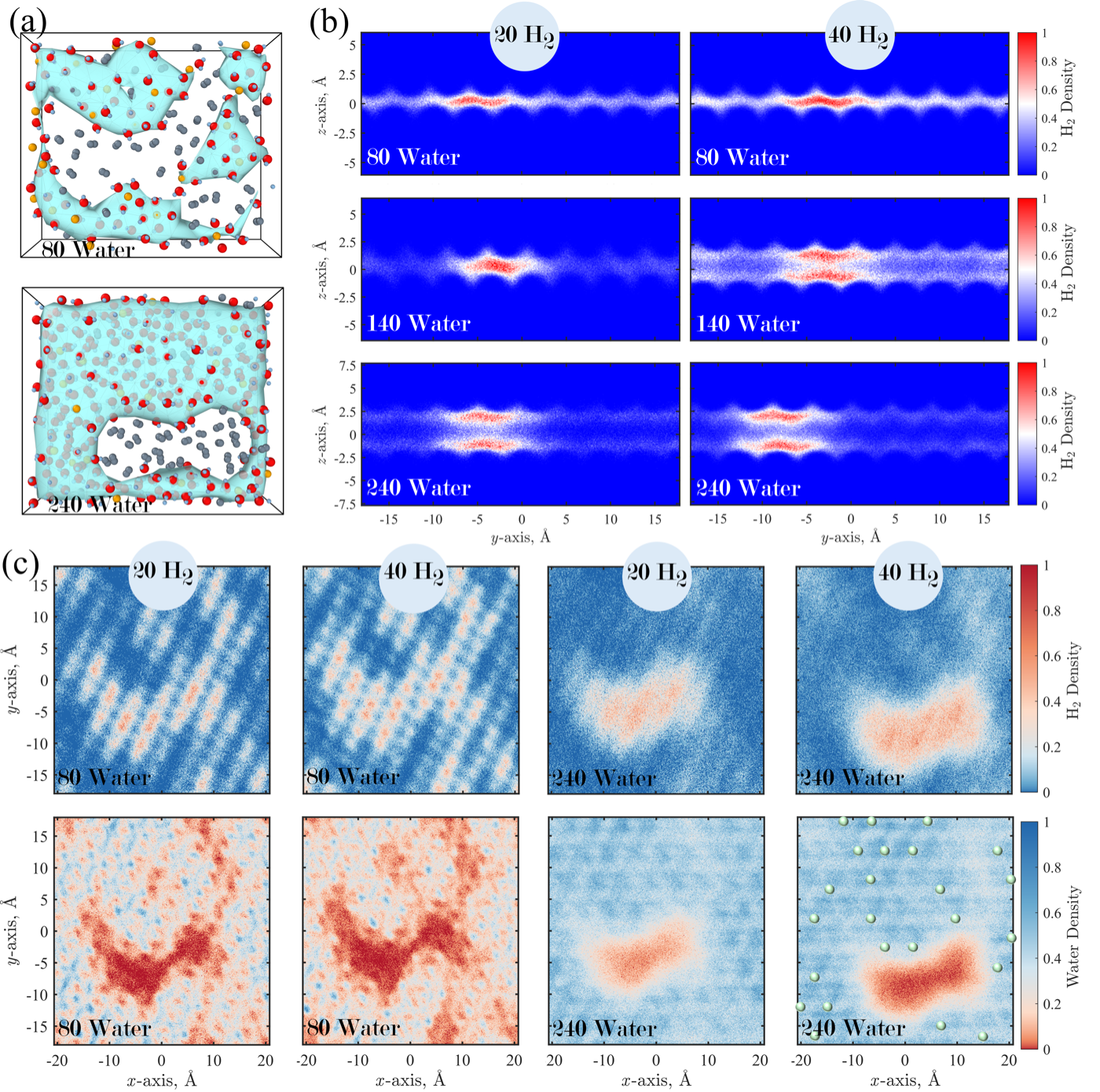}
    \caption{(a) 3D visualization of confined fluid between two Mt slabs (not shown, positioned above and below the presented layer) over the \textit{xy}-plane, comparing systems with one (80 water, 2.5 \ce{H2O}/UC) and two (240 water, 7.5 \ce{H2O}/UC) hydration layers. \ce{H2} is shown as gray van der Waals (vdW) spheres, water oxygen atoms are red, hydrogen atoms are white, and the water surface is rendered in transparent cyan. (b) 2D density maps of \ce{H2} projected onto the \textit{yz}-plane for different water and \ce{H2} contents, showing \ce{H2} plume formation. (c) 2D density maps in the \textit{xy}-plane of \ce{H2} (top row) and water (bottom row) for the different systems, illustrating the spatial distributions of the components at varying water and \ce{H2} contents.}
    \label{fig:2D-density}
\end{figure}

An important interaction mechanism arises from the steric effects of \ce{H2}, together with its gas-like behavior (not dissolved in water) and preferential distribution (Figure ~\ref{fig:2D-density}a), which tend to compress water molecules more closely together (see Figure S4, SI). This spatial confinement promotes H-bond formation among water molecules, thereby accelerating the organization of the water molecular structure. As shown in Figure~\ref{fig:Swelling-Energy-Hbonds}d, \ce{H2} enhances the water network by increasing H-bonding between water molecules (Hw-Ow), particularly at low water content. However, the transition to the 2W hydration state disrupts this network, as the formation of a second hydration layer, driven by the overall volume increase—allows water molecules to move further apart. These results also indicate that, in the 2W system, the presence of \ce{H2} has a negligible effect on Hw-Ow H-bonding. Note that as expected, Hw-Ow H-bonds in nanopores are significantly lower than in bulk ($\sim$ 3.5 H-bonds per water molecule in bulk). In addition, increasing water generally reduces water–Mt H-bonds. 

Indeed, the low solubility of \ce{H2} in water impacts its spatial distribution, particularly when its concentration exceeds the solubility limit (0.08 mol/kg, which is 1.44 $\times 10^{-3}$ \ce{H2} per 1 \ce{H2O}) \cite{ho2023low,van2023interfacial,rahbari2019solubility}. Figure~\ref{fig:2D-density}b,c present side and top views of the 2D density distribution of \ce{H2} under different hydration states. At water content below 1W hydration state, \ce{H2} appears to be uniformly distributed along the clay surface. As the water content increases and the system transitions toward a 1W and 2W hydration state (Figure \ref{fig:2D-density}b, system 80 water (2.5 \ce{H2O}/UC) and 140 water (4.375 \ce{H2O}/UC)), \ce{H2} begins to aggregate into gaseous domains, forming distinct feature of plumes. A closer examination of the side view reveals that, in the 1W state, these \ce{H2}-rich clusters are confined by surrounding water molecules, which limits their lateral mobility within the \textit{xy}-plane. Similarly in the 2W system, although this confinement is relaxed and \ce{H2} localize within each hydration layer, they are still connected as a single cohesive plume without any interruption from water molecules (Figure~\ref{fig:2D-density}c, system 240 water (7.5 \ce{H2O}/UC), 20 and 40 \ce{H2}). It should be noted that \ce{H2} nano-bubbles in bulk tend to form circular, symmetric shapes because this geometry minimizes the surface area for a given volume, thereby reducing surface energy according to the Laplace pressure principle \cite{yen2022analysis}. In contrast, our findings show that under Mt nano-confined environments with interlayer distances below 2~nm, \ce{H2} molecules form asymmetric bubble morphologies. These morphologies largely driven by the locations of isomorphic substitutions in the Mt’s octahedral sheet, O-sheet, (Figure~\ref{fig:2D-density}c shown over the water 2D density map by light green), where the arising negative charge due to Mg$^{2+}$ substitution for Al$^{3+}$ defines the local water–cation arrangement, thereby influencing the shape of the gas domain. 

\begin{figure}[H]
    \centering
    \includegraphics[width=0.95\linewidth]{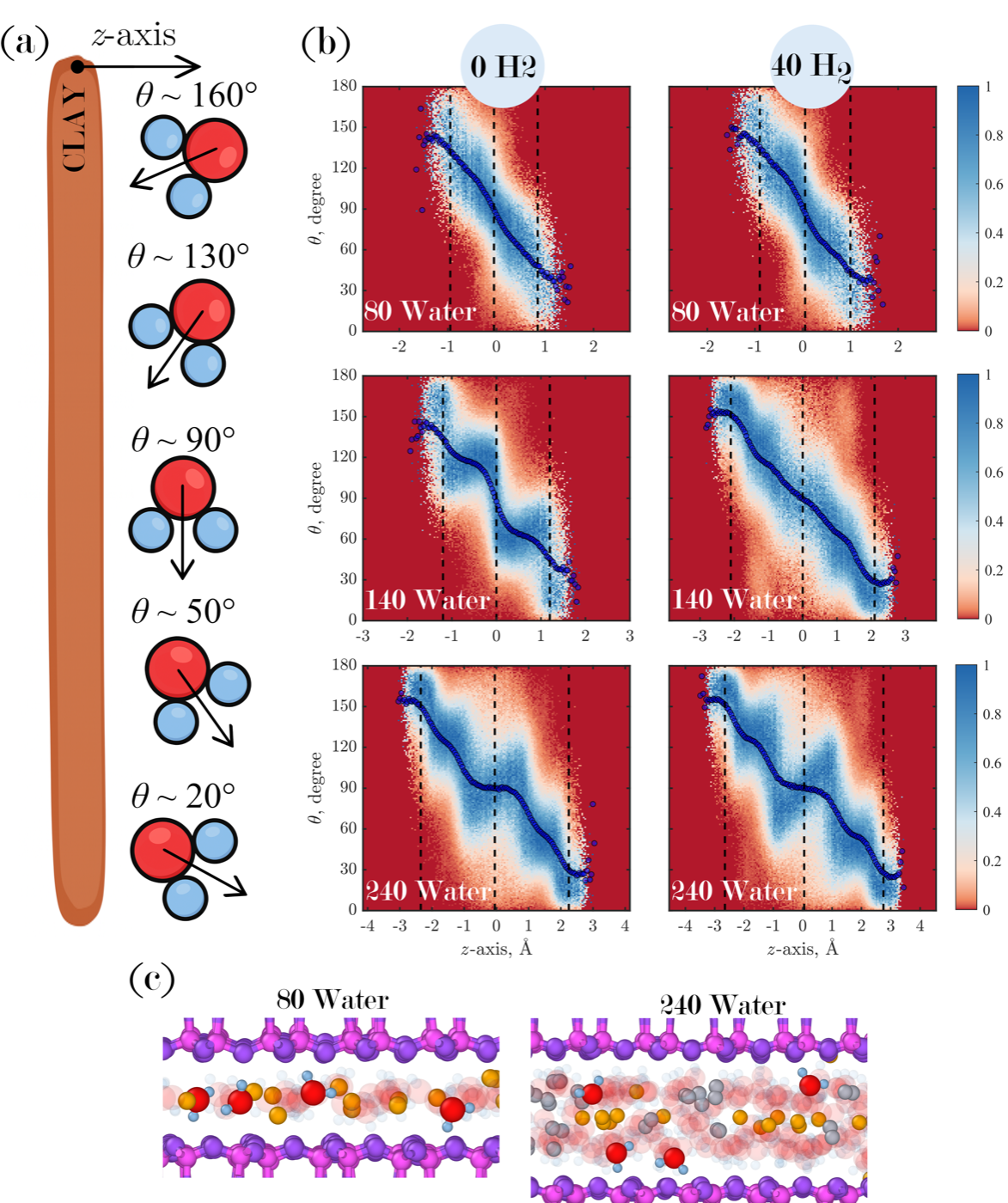}
    \caption{(a) Schematic diagram of water orientation, defined as the angle between the water dipole moment and the \textit{z}-direction (normal to the clay slab, shown in brown), (b) population of the angle formed in the interlayer along the \textit{z}-direction, from one top of the clay slab to the other. The two dashed lines on either side indicate the peaks of the water number density in the interlayer, and the blue line represents the average orientation population, and (c) illustration of water orientation in two different systems: 80 water molecules (2.5 \ce{H2O}/UC) without \ce{H2}, and 240 water molecules (7.5 \ce{H2O}/UC) with 40 \ce{H2} molecules.}
    \label{fig:orientation}
\end{figure}

Furthermore, we investigated the orientation distribution of water molecules near the clay surface (Figure~\ref{fig:orientation}a) and the corresponding average orientation population profiles (Figure~\ref{fig:orientation}b). In the absence of \ce{H2} and within the 1W hydration layer, water molecules exhibit strong orientational ordering near the clay surface, characterized by a distribution range of $110^\circ$ to $160^\circ$ and distinct average angular peaks centered around $135^\circ$ (and $45^\circ$ on the opposing surface). These orientations correspond to water molecules approaching the surface with one of their hydrogen atoms pointing toward it. As the system approaches the transition region and the formation of the 2W hydration layer, water molecules become increasingly aligned with the surface normal, with the orientation angle range progressively shifting toward $\sim130^\circ$ to $\sim170^\circ$. This indicates configurations in which both hydrogen atoms face the clay slab. Overall, this trend suggests stronger dipole alignment of water molecules toward the clay surface as hydration increases (Figure~\ref{fig:orientation}c).

The introduction of \ce{H2} molecules has no tangible impact on the orientation of water near the surface when stable 1W and 2W hydration layers are formed. In contrast, as shown in Figure~\ref{fig:orientation}b, increasing the number of \ce{H2} molecules gradually shifts the average orientation angle of water molecules near the surface in the transition region (140 water molecules in the system), with the average peak moving from $\sim140^\circ$ to $\sim150^\circ$. This shift indicates that the occupation of the interlayer midplane by \ce{H2} molecules pushes water molecules closer together, as explained earlier, forcing them to adopt configurations in which both hydrogen atoms are oriented toward the clay surface.

In summary, \ce{H2}  accelerates the formation of stable water layers in both 1W and 2W configurations at lower water contents, promoting clay swelling during anhydrous conditions and hydration state transitions (Figure~\ref{fig:Swelling-Energy-Hbonds}b). This deepens the hydration energy minimum but narrows the stable hydration range, indicating reduced structural resilience (Figure \ref{fig:Swelling-Energy-Hbonds}c). Indeed, \ce{H2} repositions Na$^+$ ions, weakening clay–cation interactions in anhydrous clay and enhancing hydration in the 2W state (Figure~\ref{fig:1D-density}). The steric effects of \ce{H2} promote H-bonding at low water content but has little effect in 2W systems (Figure~\ref{fig:Swelling-Energy-Hbonds}d), while higher water content promotes water localization within the interlayer rather than at the Mt surface.

In these nano-confinements, \ce{H2} forms asymmetric gas plumes, influenced by O-sheet clay substitutions (Figure~\ref{fig:2D-density}). \ce{H2} disrupts water orientation in the transition region without a considerable impact on the 1W and 2W states (Figure~\ref{fig:orientation}). It should be noted that T-sheet substitutions are relatively uncommon in Mt. In contrast, O-sheet substitutions impose a weaker electrostatic influence on interlayer ions. As a result, the spatial distribution of interlayer Na$^{+}$ is more strongly dictated by T-sheet substitutions, which can in turn enhance the development of plume-like ion arrangements.

\newpage
\subsection{Elastic Response under Hydration and \ce{H2}}

\begin{figure}[H]
    \centering
    \includegraphics[width=1\linewidth]{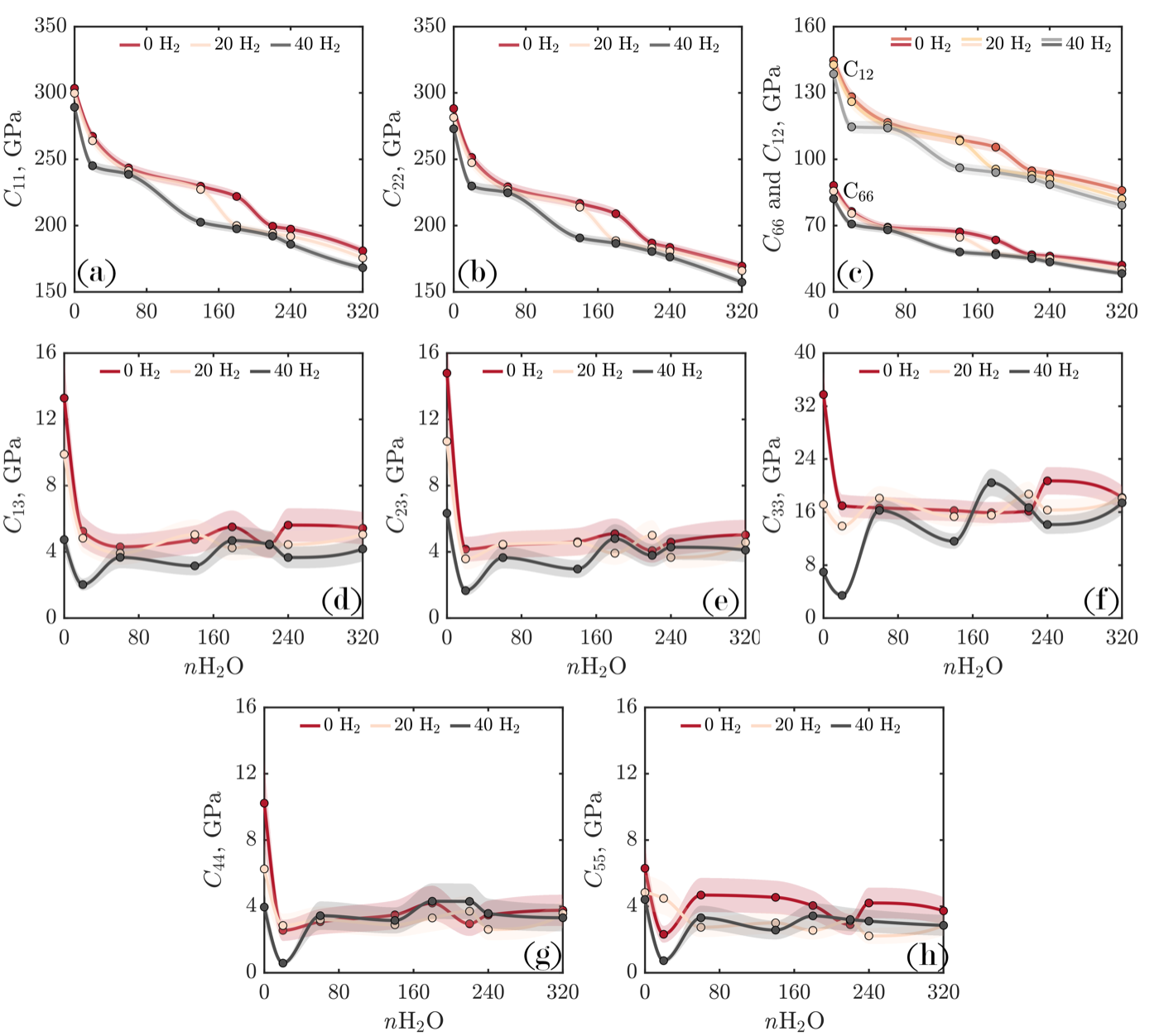}    
    \caption{Directional elastic stiffness coefficients of Mt as a function of interlayer water contents (0, 20, 60, 140, 180, 220, 240, and 320 molecules)  and \ce{H2} contents (0,20, and 40 molecules): (a–c) in-plane stiffness coefficients (\textit{C$_{11}$}, \textit{C$_{22}$}, \textit{C$_{12}$}, and \textit{C$_{66}$}); (d–h) out-of-plane stiffness coefficients (\textit{C$_{13}$}, \textit{C$_{23}$}, \textit{C$_{33}$}, \textit{C$_{44}$}, and \textit{C$_{55}$}). The shaded area indicates the mean value $\pm$ standard deviation from repeated simulations.}
    \label{fig: Mechanical-Fig5}
\end{figure}

Figure~\ref{fig: Mechanical-Fig5} shows all elastic stiffness coefficients of Na-Mt with variable interlayer water and \ce{H2} content, reflecting the anisotropic mechanical behavior arising from the stiff clay basal planes and the mechanically soft interlayer. The in-plane coefficients (\textit{C$_{11}$}, \textit{C$_{22}$}, \textit{C$_{12}$}, and \textit{C$_{66}$}) (Figure~\ref{fig: Mechanical-Fig5}a-c) describe deformation within the Mt crystal structure, while the out-of-plane coefficients (\textit{C$_{13}$}, \textit{C$_{23}$}, \textit{C$_{33}$}, \textit{C$_{44}$}, and \textit{C$_{55}$}) (Figure~\ref{fig: Mechanical-Fig5}d-h) characterize the mechanical response involving the interlayer space.

As expected from the layered silicate structure, the in-plane coefficients are significantly higher than the out-of-plane coefficients, with the \textit{C$_{11}$/C$_{55}$} ratio of  around 45 for the dry system, underscoring the strong intralayer covalent/ionic bonding and the relatively weak van der Waals and H-bonding in the interlayer. This pronounced anisotropy is consistent with findings from previous molecular simulations and experimental studies \cite{mazo2008molecular, ortega2007effect, chen2006elastic}. Among the in-plane coefficients, \textit{C$_{11}$} is the highest, followed by \textit{C$_{22}$}, \textit{C$_{12}$}, and \textit{C$_{66}$}, the last of which reflects shear stiffness within the basal plane and is the lowest among the group. This is consistent with trends reported in both experimental and first-principles investigations \cite{teich2012molecular, militzer2011first}. Notably, \textit{C$_{11}$} and \textit{C$_{22}$} differ by $\sim$ 6\%, which falls within the range (2–10\%) reported for other phyllosilicates. \cite{mcneil1993elastic, zartman2010nanoscale}

Increasing interlayer water content reduces all in-plane coefficients, regardless of the amount of \ce{H2} present, particularly pronounced at low hydration levels. For example, adding 20 water molecules to the dry system decreases in-plane coefficients by $\sim$ 12\% on average. In contrast, \ce{H2} alone leads to smaller reductions from 1.5 to 9\% for 20 to 40 \ce{H2} molecule systems, with the magnitude of this effect varying depending on the hydration level. This trend has been attributed to an increase in basal spacing and has been previously reported for \ce{CO2}-containing clay swelling systems\cite{zhang2015interplay,zhang2017evolutions}. 

In contrast, the out-of-plane coefficients show greater sensitivity to shear and normal stresses perpendicular to the layers than to basal spacing, and drop significantly with the addition of initial water or \ce{H2}. For example, the addition of 20 water molecules reduces out-of-plane coefficients by 50 to 75\%. The entry of the first water and \ce{H2} molecules into the interlayer space of the dry system results in a much greater reduction in out-of-plane coefficients than in in-plane ones. Interestingly, at higher hydration levels (60 water molecules in the system, 1.875 \ce{H2O}/UC), a partial recovery of certain out-of-plane coefficients is sometimes observed. 
This non-monotonic behavior has previously been attributed to the formation of structured water networks and an average increase in H-bonding between Mt and water (see Figure~\ref{fig:Swelling-Energy-Hbonds}d).\cite{zhang2017evolutions} As the number of water molecules increases beyond 60, the out-of-plane coefficients tend to fluctuate around a steady value, without showing a specific trend or dependence on the \ce{H2} concentration, but exhibiting heightened fluctuations at higher \ce{H2} levels. This triphasic behavior, initial softening, partial stiffening, and subsequent oscillatory response, has been described as a characteristic feature of water-induced mechanical transitions in swelling clays and remains applicable to the \ce{H2}-containing system \cite{ebrahimi2012nanoscale, carrier2014elastic}.

Furthermore, the bulk modulus ($K$, resistance to volumetric change), shear modulus ($G$, resistance to shape change), and Poisson’s ratio ($\nu$, ratio of lateral to axial strain) of Mt systems as a function of interlayer water and \ce{H2} content are presented in Figure~S5, SI. The variations in $K$ and $G$ as a function of water and \ce{H2} content reflect a combination of the stiffness trends of in-plane and out-of-plane coefficients, described above. Overall, increasing water and \ce{H2} content leads to a decrease in both $K$ and $G$, although some exceptions occur due to the stiffness contribution of out-of-plane coefficients. This decrease is most pronounced when moving from the dry system to systems with low water and \ce{H2} content. In contrast, $\nu$ does not follow a systematic trend with water or \ce{H2} content, but instead shows irregular fluctuations.

\subsection{Effect of Induced Swelling on Elastic Behavior} 

Water and \ce{H2} have a paradoxical influence on the in-plane stiffness coefficients, exhibiting both reinforcing and weakening effects. Given the positions of Si and Al in tetrahedral and octahedral sites, respectively, analyzing interactions between them and surrounding oxygens can reveal key factors on the structural stability of Mt. Figure S6, SI, illustrates the RDFs of Al$\cdots$O and Si$\cdots$O bonded pairs at varying levels of water and \ce{H2} content. Across different systems, the positions of the first peaks remain unchanged (1.93 \AA{} for Al$\cdots$Ob, 1.93 \AA{} for Al$\cdots$Oh, and 1.58 \AA{} for Si$\cdots$Ob). However, increasing the water and \ce{H2} content amplifies the first peak in all three cases, indicating strengthened interactions in these bonded pairs \cite{li2024understanding,wei2022effect}. As a result, higher water and \ce{H2} content marginally enhance the strength of the intralayer clay structure. These changes in intralayer interactions may be related to variations in the attractive and repulsive forces between the Mt layers and the interlayer, resulting from changes in interlayer content \cite{li2020internal}. Despite this local strengthening, as discussed earlier, the in-plane coefficients decrease as interlayer water and \ce{H2} content increase, expanding the interlayer distance and enlarging the cross-sectional area ($A$) over which the force ($F$) is distributed. This geometric effect (\textit{i.e.}, the impact of induced swelling) lowers the stress ($\sigma = F/A$) for a given force. Consequently, the in-plane coefficient, proportional to the stress–strain ratio, decreases with increasing basal spacing. This geometrically induced softening outweighs local structural reinforcement, making the coefficients depend more on basal spacing than on bond strength, highlighting the interplay between structural and geometric factors in hydrated Mt elasticity.


The geometric mechanism associated with variations in basal spacing does not affect the out-of-plane coefficients, since cross-sectional area \textit{A} for shear or normal (\textit{z}-direction) forces remains constant. However, water weakens these coefficients by replacing strong Na$^+$–Mt electrostatic interactions with weaker H-bonds and by forming hydration shells that displace Na$^+$ from the Mt surface (Figure~\ref{fig:1D-density}). In contrast, non-polar \ce{H2} does not form H-bonds with the Mt surface nor hydrate Na$^+$ ions; instead, it disrupts ionic bridges without forming compensatory bonds. In the absence of interlayer water or \ce{H2}, Na$^+$ ions become locked between the Mt layers, often located next to isomorphic substitutions. However, in the presence of \ce{H2}, the high mobility of the gas molecules alters the positions of Na$^+$ ions. The initial incorporation of either water or \ce{H2} acts as a lubricant, reducing interlayer friction and promoting slippage between Mt layers, which in turn, lowers the system's shear strength. Importantly, this molecular inclusion significantly weakens Na$^+$–Mt interactions, with the binding interaction energy dropping from $-12769\pm10$ kJ/mol in the dry system to $-9574\pm14$ kJ/mol (25\%) and $-10605\pm11$ kJ/mol (17\%) following the addition of 20 water molecules and 20 \ce{H2} molecules, respectively. This sharp reduction at the early stage highlights the extreme sensitivity of interlayer cohesion to even minimal amounts of water or \ce{H2}.

\subsection{Mechanical Failure Response under Hydration and \ce{H2}} 

\begin{figure}[H]
    \centering
    \includegraphics[width=1\linewidth]{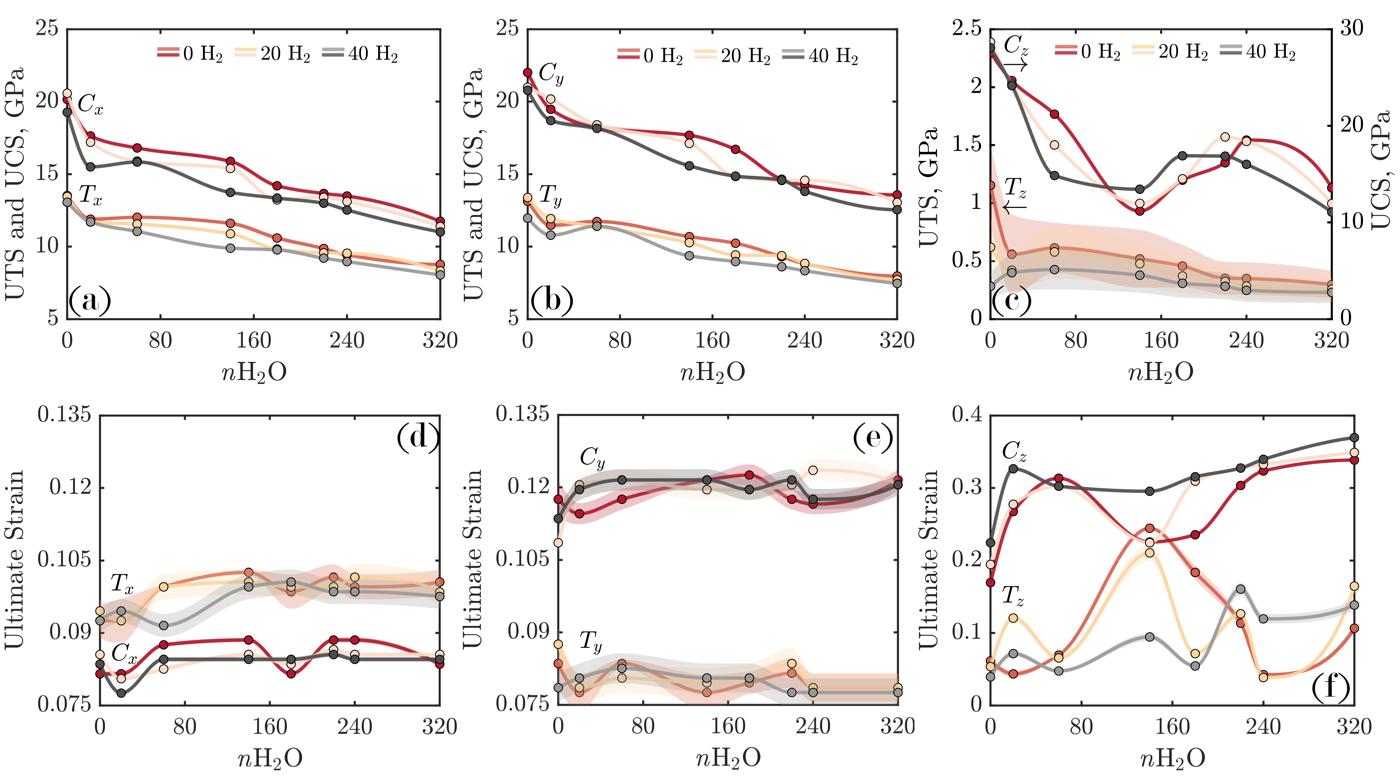}
    \caption{\textit{UTS} and \textit{UCS} (a–c) and ultimate strain (d–f) as a functions of interlayer water contents (0, 20, 60, 140, 180, 220, 240, and 320 molecules) and \ce{H2} contents (0,20, and 40 molecules) for systems loaded along the (a,d) \textit{x}-, (b,e) \textit{y}-, and (c,f) \textit{z}-directions. The shaded area indicates the mean value $\pm$ standard deviation from repeated simulations.}
    \label{fig: UTS-UCS}
\end{figure}

The ultimate tensile strength (\textit{UTS}) and ultimate compressive strength (\textit{UCS}) are the maximum stresses a material can withstand under tension and compression, respectively, before failure. Accordingly, Figures~S7 and S8, SI, were used to determine the \textit{UTS}, \textit{UCS}, and ultimate tensile and compressive strain values. Figure~\ref{fig: UTS-UCS}a-c presents the triaxial \textit{UTS} and \textit{UCS} values for Mt systems containing varying amounts of water and \ce{H2}. Pronounced mechanical anisotropy is evident: for every system studied, the $UTS$ follows the order: $UTS_{x} > UTS_{y} > UTS_{z}$, whereas the $UCS$ follows: $UCS_{z} > UCS_{y} > UCS_{x}$. These rankings, well documented in prior work—reflect the layered structure of Mt. \cite{li2023deformation,wei2022effect,wei2023atomistic}. While two systems, 20 \ce{H2O}–20 \ce{H2} and 60 \ce{H2O}–20 \ce{H2}, deviate slightly from the expected tensile‐strength ordering, the differences fall within the numerical uncertainty of the simulations; nonetheless, they illustrate that \ce{H2} can locally perturb the anisotropy hierarchy. In all cases, the \textit{UCS} exceeds the corresponding \textit{UTS}, indicating a greater tensile vulnerability of clay, particularly along the \textit{z}-direction, where plasticity precedes fracture. This observation is consistent with the macroscopic experimental findings of Zhang \textit{et al}. \cite {zhang2019investigation,zhang2019molecular}. The dry system without \ce{H2} shows the highest \textit{UTS} in the \textit{z}-direction, while the initial addition of 20 \ce{H2} and 20 \ce{H2O} reduces \textit{UTS} by 46\% and  52\%, respectively, due to disruption of Na$^+$–Mt ionic bonds, consistent with the out-of-plane stiffness coefficients (Figure \ref{fig: UTS-UCS}c).

The ultimate tensile strain and ultimate compressive strain (\textit{i.e.,} the strains at \textit{UTS} and \textit{UCS}) quantify the deformation a system can sustain before failure. As shown in Figure \ref{fig: UTS-UCS}d-f, Mt slabs loaded in the \textit{x}-direction tolerate, on average, 18\% higher ultimate tensile strain than those loaded in the \textit{y}-direction. Under compression, \textit{y}-direction loading exhibits around 40\% higher ultimate compressive strain than \textit{x}-direction loading, reflecting greater ductility along \textit{y}. Strain fluctuations are largest along the \textit{z}-direction, consistent with the pronounced plastic response observed there (Figure \ref{fig: UTS-UCS}c). Tensile loading in the \textit{y}-direction initiates cracks faster than compression, whereas in the \textit{x}-direction, compression causes earlier failure than tension. In addition, neither \ce{H2} nor water significantly alters the ultimate strain in the \textit{x}- and \textit{y}-directions, and no consistent trend is observed across varying conditions (Figure \ref{fig: UTS-UCS}d,e). In contrast, the \textit{z}-direction shows non-monotonic variations: compressive ductility first increases, then decreases, and increases again with changing water content (Figure \ref{fig: UTS-UCS}f). Our findings align with those of Zhao \textit{et al}. \cite{zhao2023tensile}. The dry, \ce{H2}-free system fails fastest in \textit{z}-direction compression, while the most dramatic changes in ductility occur upon the first inclusion of guest molecules: introducing 20 water molecules or 20 \ce{H2} molecules enhances \textit{z}-direction ultimate compressive strain by 64\% and 16\%, respectively, relative to the dry reference. This arises from the much higher compressibility of systems containing water and \ce{H2} compared to the dry, \ce{H2}-free system.


\subsection{Mechanical Failure Behavior}

Because the clay layers lie in the \textit{xy}-plane, the in-plane \textit{UTS} and \textit{UCS} are governed by bond rupture within the Mt lattice and are determined from the elastic portions of the stress–strain curves for loading along the \textit{x}- and \textit{y}-directions, which exhibit constant slopes (Figures S7 and S8, SI). Given this linearity, \textit{UTS} and \textit{UCS} primarily depend on the corresponding elastic stiffness coefficients (\textit{C$_{11}$}, \textit{C$_{22}$}) and the ultimate tensile or compressive strain (Figure \ref{fig: UTS-UCS}d,e). Consequently, the in-plane strength trends in Figure \ref{fig: UTS-UCS}a,b reflect variations in \textit{C$_{11}$} and \textit{C$_{22}$} (Figure \ref{fig: Mechanical-Fig5}a,b) with minor deviations arising from differences in ultimate strains (Figure \ref{fig: UTS-UCS}d,e). It should be noted that, as these coefficients vary systematically with basal spacing, in-plane \textit{UTS} and \textit{UCS} are likewise governed by this relationship, with ultimate strain playing a secondary role.

\begin{figure} [H]
    \centering
    \includegraphics[width=0.95\linewidth]{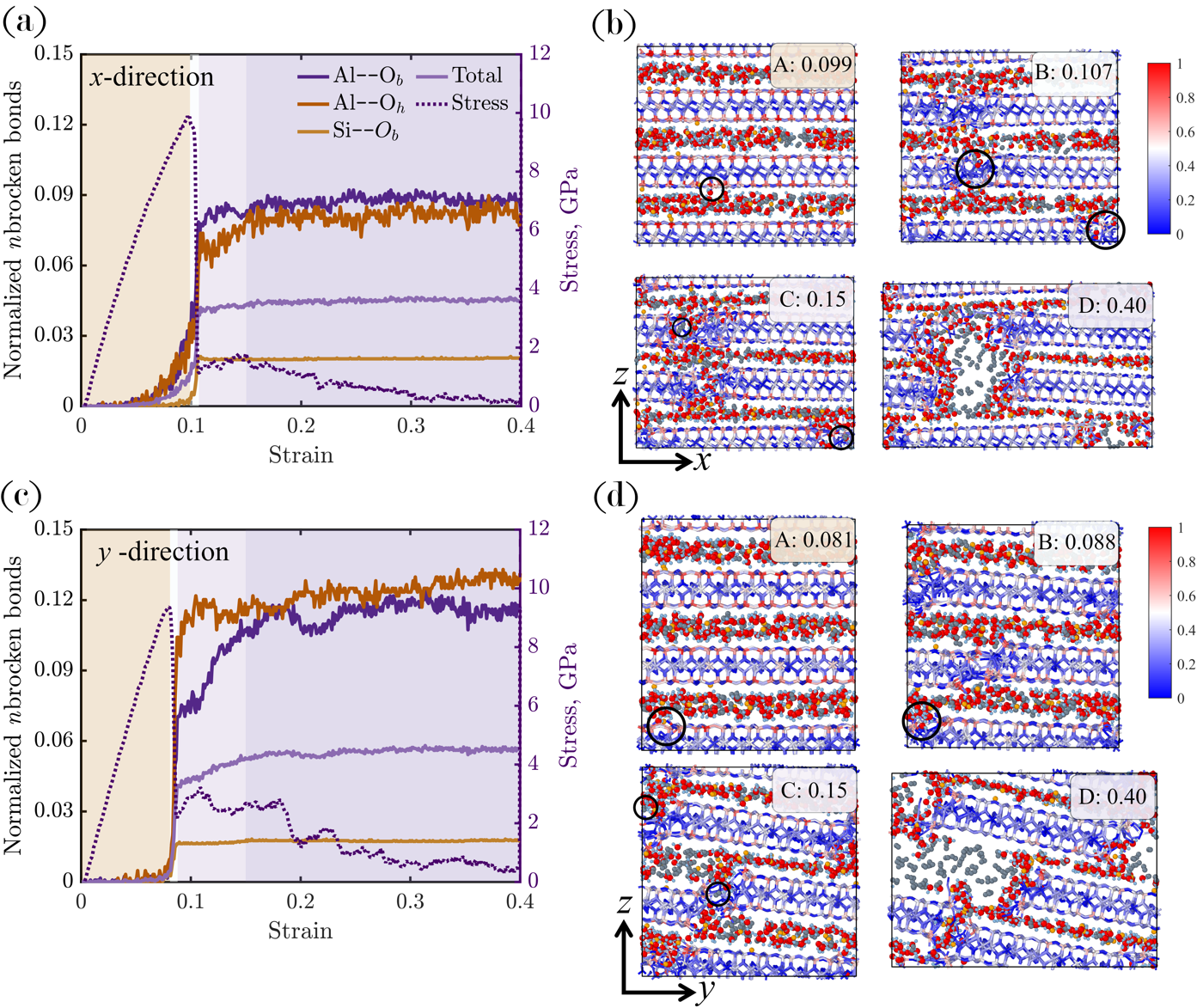}
    \caption{Normalized number of broken bonds for Al$\cdots$Ob and Al$\cdots$Oh in the O-sheet, and Si$\cdots$Ob in the T-sheet (left axis), together with the corresponding stress variation (right axis) at different strain levels for (a) \textit{x}-direction and (c) \textit{y}-direction tension in the system containing 140 water and 40 \ce{H2} molecules. Panels (b) and (d) present visualizations and normalized stress distribution in the Mt structure during tensile loading at critical strain values in the (b) \textit{x}-direction and (d) \textit{y}-direction. According to the colormap, red and blue in the Mt structure indicate high and low stress values, respectively.}
    \label{fig: visual-Tx-Ty}
\end{figure}

To further assess the role of atomic pairs in the \textit{UTS} and \textit{UCS} of the \textit{xy}-plane, we examine the structural deformation and the normalized number of broken bonds in the O- and T-sheets, focusing on Al–basal oxygen pairs (Al$\cdots$Ob), Al–hydroxyl oxygen pairs within the O-sheet (Al$\cdots$Oh), and Si–basal oxygen pairs within the T-sheet (Si$\cdots$Ob) under tension in the \textit{x}- and \textit{y}- directions. Figure \ref{fig: visual-Tx-Ty} shows the visual analysis of mechanical asymmetry alongside the normalized stress distribution within the Mt structure for a selected system with 140 water molecules (4.375 \ce{H2O}/UC) and 40 \ce{H2} molecules. Note that the virtual broken bond criteria follow Yang \textit{et al.}\cite{yang2019deformation}, using RDF cutoff radii of 1.84 \AA~ for Si$\cdots$Ob, 2.66 \AA~ for Al$\cdots$Ob, and 2.84 \AA~ for Al$\cdots$Oh, consistent with established literature values\cite{li2024understanding,wei2022effect} (see Figure~S6, SI).

Considering the critical strain points A–D (Figure~\ref{fig: visual-Tx-Ty}b,d), stage A corresponds to maximum stress (0.099 and 0.081 for the \textit{x}- and \textit{y}-directions, respectively, Figure \ref{fig: visual-Tx-Ty}a,c), marking failure. Stage B (0.107 in \textit{x}- and 0.088 in \textit{y}-directions) represents the post-failure phase, characterized by stress relaxation. With increasing strain, interlayer spacing decreases due to the Poisson effect. As shown by the renderings in Figure \ref{fig: visual-Tx-Ty}b,d, stress buildup is greater in the T-sheet than the O-sheet; however, normalized data reveal more bond fractures in the O-sheet despite its higher rupture threshold relative to the T-sheet. Under \textit{x}-direction tension, bond breakage follows Al$\cdots$Ob $>$ Al$\cdots$Oh $>$ Si$\cdots$Ob, while in the \textit{y}-direction it is Al$\cdots$Oh $>$ Al$\cdots$Ob $>$ Si$\cdots$Ob. Fractures initiate mainly in O-sheet bonds under \textit{x}-direction tension before propagating to the T-sheet, indicating stronger tetrahedral bonding. Overall, Al$\cdots$Ob and Al$\cdots$Oh breaks are more frequent in \textit{y}-direction tension, while Si$\cdots$Ob dominates under \textit{x}-direction tension. These trends reflect not only bond strength but also bond orientation, atomic configuration, and mobility: Si$\cdots$Ob bonds in the T-sheet have tetrahedral coordination, whereas Al$\cdots$Ob bonds in the O-sheet are octahedral. Larger T-sheet displacements in \textit{x}-direction tension yield higher fracture rates, consistent with the increased \textit{UTS} along this axis.
As shown in Figure \ref{fig: visual-Tx-Ty}b,d, stage A exhibits pronounced stress concentrations in the T-sheet. Following the fracture in stage B, stress localizes around crack sites within the T-sheet, where interlayer molecules apply additional pressure that promotes crack growth. Most bond breakage occurs between stages A and B, after which the system stabilizes. Under \textit{x}-direction tension, cracks are vertically aligned, in contrast to the more irregular patterns observed under \textit{y}-direction tension. Note that the Si$\cdots$Ob broken bond diagram aligns closely with the stress–strain curve shown in Figure~\ref{fig: visual-Tx-Ty}a,c (dashed line, right axis). No T-sheet bond breakage occurs during the linear stress increase, but the broken bond coincides exactly with the sudden stress drop, highlighting the T-sheet’s critical role, over the O-sheet, in tensile failure. The variations in, and contributions of, each sheet to the system’s total potential energy, shown in Figure~S9, SI, further support this.

Bond breakage in the T-sheet enables interlayer molecules to penetrate the Mt structure (circle in stage A). Water infiltrates fractures more readily than \ce{H2} due to stronger clay interactions (in the areas enclosed by black circles in stages B). With continued deformation, \ce{H2} also enters alongside water, aided by its smaller size and higher diffusivity (circles in stage C). This co-penetration weakens interlayer bonds, reduces stress, and separates clay platelets. Under \textit{y}-direction tension, platelets tilt and disconnect nanopores, while \textit{x}-direction tension preserves or strengthens connectivity. At maximum \textit{x}-strain (0.4, stage D), cracks widen to uptake water, and \ce{H2} circulates through central channels—behavior absent under \textit{y}-direction tension, where \ce{H2} remains confined.

\newpage

\begin{figure}[H]
    \centering
    \includegraphics[width=0.95\linewidth]{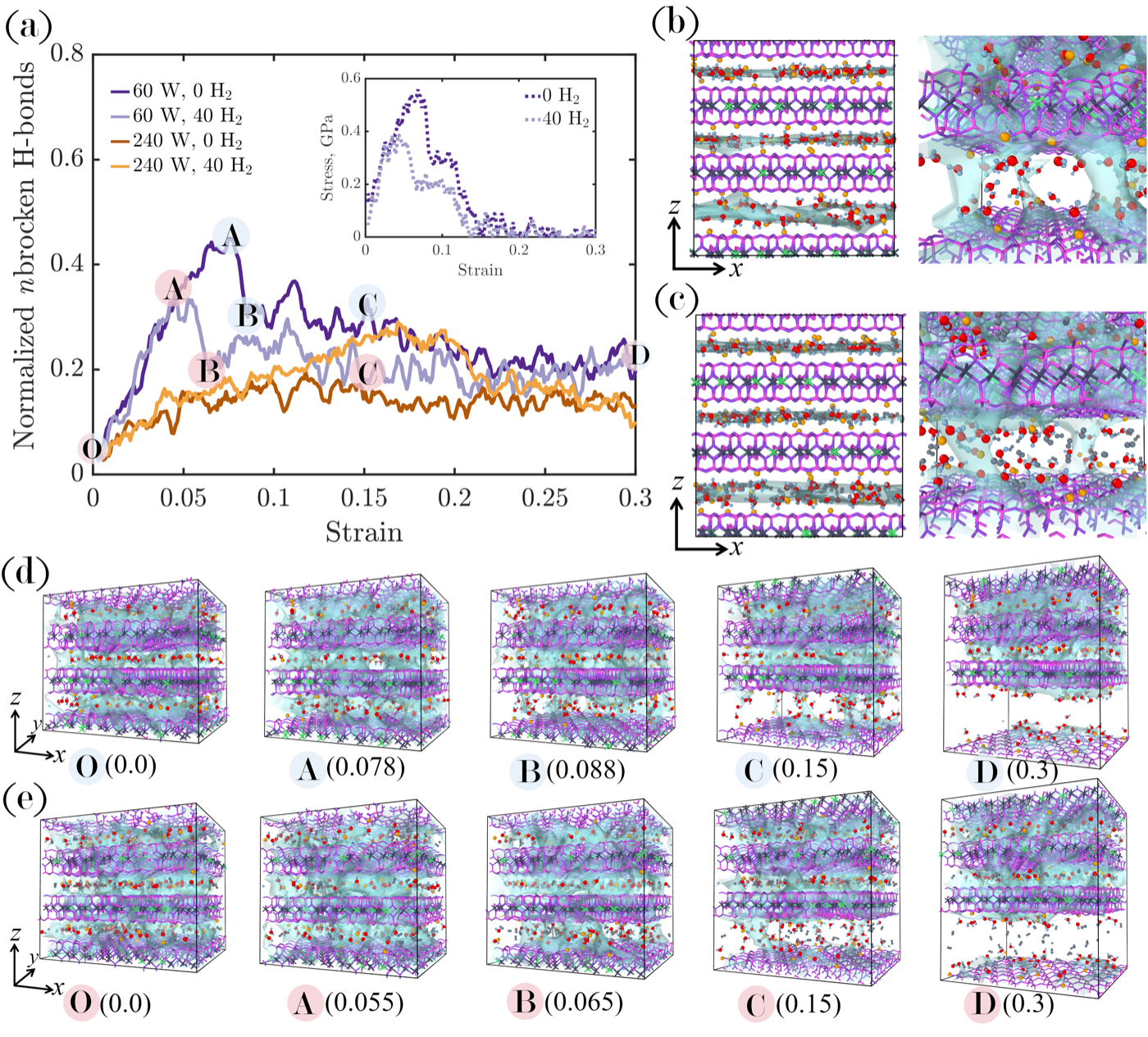}
    \caption{(a) Normalized number of broken H-bonds between Mt and interlayer water during \textit{z}-direction tension for systems containing 60 and 240 water molecules, both without \ce{H2} and with 40 \ce{H2} molecules (inset shows the stress response for the 60-water system). (b, c) Side and close-up views at points B and C for the 60-water system without \ce{H2} (b) and with 40 \ce{H2} molecules (c). (d, e) Tensile deformation in the \textit{z}-direction at the indicated points for the 60-water system without \ce{H2} (d) and with 40 \ce{H2} molecules (e).}
    \label{fig: Broken-Hbond}
\end{figure}

As shown in Figure \ref{fig: UTS-UCS}c, increasing water and \ce{H2} contents reduce the \textit{UTS} in the out-of-plane direction. At higher water contents, the effect of \ce{H2} becomes negligible and falls within the uncertainty range for 2W systems. Due to the surface charge of Mt, water–Mt H-bonds (Hw–Ob) are stronger than water–water bonds (Hw–Ow)\cite{rastogi2011hydrogen}, allowing Hw–Ob bonds to better resist separation under out-of-plane loading. To capture these effects in hydrated systems, we analyze the normalized number of broken Ob–Hw bonds. Figure~\ref{fig: Broken-Hbond}a shows this metric for systems with 60 (1.875 \ce{H2O}/UC) and 240 (7.5 \ce{H2O}/UC) water molecules, with and without 40 \ce{H2}, from O (no strain) to D (final strain), along with the stress–strain curve (top right) for the 60-water system. O marks the unstrained state, A the peak in broken bonds, B the subsequent drop, and C and D strains of 0.15 and 0.3. Figure~\ref{fig: Broken-Hbond}b,c show the side view at B (onset of bilayer water) and the close-up at C (nanoscale liquid-bridge stage) for 60-water systems without and with 40 \ce{H2} molecules, respectively. Figure~\ref{fig: Broken-Hbond}d,e illustrate tensile-loading trajectories at the five labeled points in the out-of-plane direction.

The disruption process begins with a sharp rise (O $\rightarrow$ A) in broken H-bonds, during which water inertia resists layer separation, increasing the water–surface oxygen distance, misaligning orientations, and destabilizing Na$^+$ hydration shells. This is followed by a sharp decrease (A $\rightarrow$ B) as water migrates vertically to form a bilayer (see Figure~\ref{fig: Broken-Hbond}b,c, \textit{x-z} view), reducing the average surface distance and reforming broken bonds. Finally, the process enters a plateau (B $\rightarrow$ D), where thin water films line the Mt surfaces and nanoscale liquid bridges form, resisting separation via surface tension and capillary suction forces (Figure~\ref{fig: Broken-Hbond}c, perspective view). These stages are more pronounced in systems with lower water content, as a greater number of H-bonds are broken during the O $\rightarrow$ A stage, which means the three phases are more distinct in systems containing 60 water molecules than in those containing 240. In \ce{H2}-containing systems, once two thin water films coat the Mt sheets (point C), \ce{H2} molecules accumulate in the central interlayer region (Figure~\ref{fig: Broken-Hbond}c, close-up view). By comparison Figure \ref{fig: Broken-Hbond}b,c, close-up, it is clear that, acting as a non-wetting, nonpolar species unable to form H-bonds, \ce{H2} forms a distinct phase that disrupts liquid bridges, weakens interfacial forces, and lowers surface tension. As strain increases and the interlayer spacing expands, these bridges break, reducing stress. At point D, the Mt sheets are fully separated, with thin water layers covering them and \ce{H2} concentrated at the nanopore center (Figure~\ref{fig: Broken-Hbond}e).

The correlation between the stress–strain curve and the normalized number of broken H-bond diagram (Figure~\ref{fig: Broken-Hbond}a) underscores the key role of H-bond disruption between Mt and water in controlling the system’s mechanical behavior along the \textit{z}-direction. H-bonds at the Mt interface act like microscopic springs, resisting vertical separation, and the applied stress primarily works to overcome them. Increasing the water content from 60 to 240 molecules significantly reduces the number of broken H-bonds, thereby lowering the \textit{UTS} in the \textit{z}-direction. As discussed above, increasing water reduces the ratio of the number of H-bonds formed between water and Mt molecules to the number of H-bonds formed between water molecules (Figure~\ref{fig:Swelling-Energy-Hbonds}d), weakening surface adhesion and lowering the \textit{UTS} in the \textit{z}-direction (Figure~\ref{fig: UTS-UCS}c). This observation is consistent with the findings of Du \textit{et al.}\cite{du2024influence} on shear friction. Comparing systems without \ce{H2} to those with 40 \ce{H2} molecules shows a slight reduction in broken H-bonds at low water content but little effect at higher content, which is why \ce{H2}’s influence is greater at low water contents, reducing \textit{UTS} in the \textit{z}-direction.

\newpage
\section{Conclusions}
This study provides critical insights into \ce{H2}-induced swelling and mechanical degradation in smectite clays, advancing the understanding of caprock integrity in subsurface \ce{H2} storage. By quantifying \ce{H2} effects on basal spacing, hydration energetics, and elastic–failure behavior, the work underscores the risks of diminished structural resilience and elevated leakage potential. The primary finding of this study are:
\begin{itemize}

\item In the presence of \ce{H2}, Mt swelling is markedly accelerated by altering the progression of hydration-state transitions. Under anhydrous conditions, \ce{H2} expands the basal spacing from 9.7~\AA\ to 10.4~\AA\ (with addition of 40 interlayer \ce{H2} molecules), accommodating gas molecules within the interlayer. In hydrated systems, \ce{H2} reduces the amount of water required to trigger 1W and 2W transitions, thereby promoting earlier onset of swelling. While \ce{H2} slightly deepens the hydration energy minimum, $–53.77$ kJ/mol, compared to  $-41.5 $ kJ/mol of internal energy of adopted water model, it also narrows the stability window of the crystalline swelling regime (1.875–3.75 \ce{H2O}/UC with \ce{H2} vs. 1.875–5.0 \ce{H2O}/UC without \ce{H2}), indicating diminished structural resilience under variable hydration states.
\item Above the solubility limit of \ce{H2}, gas-like plumes form, confined by water in the 1W state and evolving into cohesive clusters in the 2W state, thereby affecting interlayer cohesion and swelling dynamics. The asymmetric morphologies of these plumes, strongly influenced by isomorphic substitution sites in the octahedral sheet, reflect the interplay between local charge heterogeneity and nanoscale confinement. Structurally, these \ce{H2} plumes enhance H-bonding within the interlayer water network at low hydration but disrupt it in the 2W state.

\item Molecular interactions within the Mt interlayer show that \ce{H2} repositions \ce{Na+} ions and modulates water structure. In anhydrous Mt, \ce{H2} shifts \ce{Na+} closer to the clay surfaces, weakening ionic bonding, while in the 2W state it situates between the clay and cations, enhancing hydration and interlayer expansion. At low hydration, \ce{H2} disrupts water orientation, shifting dipole angles such that water molecules reorient with hydrogen atoms facing the slab surface, but its effect is minimal in stable 1W and 2W states due to strong water–water interactions.

\item Mechanically, increasing the water and \ce{H2} content in the Mt system reduces \textit{UTS} and  \textit{UCS} in the \textit{x}- and \textit{y}-directions, along with lowering the in-plane stiffness coefficients, primarily due to the expansion of the basal spacing. The \textit{z}-direction $UTS$ and the out-of-plane stiffness coefficients are even more strongly affected by 50 to 75\% for out-of-plane stiffness coefficients with 20 \ce{H2}, because the initial introduction of water and \ce{H2} disrupts Na$^+$–Mt interactions. Both bulk modulus, $K$, and shear modulus, $G$, decrease with increasing water and \ce{H2}, driven by the combined effects of basal spacing expansion and disruption of the Na$^+$–Mt interface, whereas changes in Poisson’s ratio, $\nu$, does not follow a specific trend and fluctuates with water and \ce{H2} content. These findings underscore the mechanical anisotropy of Mt and its critical role in modulating the effects of water and \ce{H2}.

\item H-bonding at the Mt interface in non-dry systems governs the $UTS$ in the \textit{z}-direction. Additionally, \ce{H2} disrupts nanoscale liquid bridges, facilitating the separation of Mt sheets. Under \textit{x}- and \textit{y}-directional tension, stress localizes in the T-sheet while bond breakage is more pronounced in the O-sheet. The stronger nanopore connectivity formed under \textit{x}-direction tension is likely to increase the probability of \ce{H2} leakage, in contrast to the weaker connectivity caused by platelet tilting in the \textit{y}-direction, highlighting anisotropic failure modes that present critical vulnerabilities for \ce{H2} storage in clay-rich caprock.

\end{itemize}

These findings challenge the applicability of \ce{CO2}-based models for \ce{H2} storage, highlighting the need for tailored strategies to mitigate geo-mechanical instabilities in clay-rich caprock. The molecular mechanisms identified here, supported by comprehensive simulations, provide a foundation for optimizing storage site selection and engineering solutions to improve long-term caprock sealing efficacy in geological \ce{H2} storage systems. From a practical standpoint, this highlights the importance of characterizing clay mineralogy at the interlayer scale, for example, distinguishing between expandable smectite-rich systems and more stable illitic or chloritic phases, since swelling and mechanical softening are strongly governed by interlayer fluid content. Incorporating hydration state and cation composition, in addition to clay mineral type, into site-screening criteria would enhance predictions of caprock integrity under \ce{H2} storage conditions.
Meanwhile, it should be noted that our study considered an ideal Mt structure in terms of isomorphic substitutions, although such substitutions can significantly influence interlayer arrangements. Other clay minerals are therefore expected to behave differently depending on their specific structures. In addition, we focused on Na$^{+}$ as the interlayer cation, whereas other common cations such as K$^{+}$ and Ca$^{2+}$ may alter the final arrangements due to their different charge densities and tendencies to form bonds with water molecules and basal surfaces. Further investigation is required to evaluate the impact of these factors.

\newpage
\section{CRediT authorship contribution statement}
\textbf{Mehdi Ghasemi}: Conceptualization, Formal analysis, Investigation, Methodology, Software, Visualization, Data curation, Supervision, Writing $-$ original draft, Writing $-$ review \& editing. \textbf{Mohamad Ali Ghafari}: Formal analysis, Methodology, Software, Validation, Data curation, Writing $-$ original draft. \textbf{Masoud Babaei}: Conceptualization, Supervision, Writing $-$ review \& editing. \textbf{Valentina Erastova}: Conceptualization, Investigation, Supervision, Writing $-$ review \& editing. 


\begin{suppinfo}
Supporting Information is available under the title 
\textit{``Molecular Insights into Caprock Integrity of Subsurface Hydrogen Storage: Perspective on Hydrogen-Induced Swelling and Mechanical Response.''} 
The Supporting Information includes: (1) methodology for computing elastic constants, 
(2) detailed description of analyses, 
(3) validation of the simulation procedure, and 
(4) additional results and discussion.\end{suppinfo}


\begin{acknowledgement}
Mehdi Ghasemi acknowledges support from the Clay Minerals Group of the Mineralogical Society (UK and Ireland) and the University of Manchester Dean’s Doctoral Scholarship. The authors also thank Research IT and the Computational Shared Facility at the University of Manchester for their support.
\end{acknowledgement}


\newpage

\bibliography{Ref}

\end{document}